\documentclass{article}
\usepackage{spconf,amsmath,graphicx,hyperref}
\usepackage{cite}
\usepackage{algorithm}
\usepackage{amssymb} 
\usepackage{algpseudocode}
\usepackage{enumitem}
\setlist[enumerate]{itemsep=0.25ex, topsep=0.3ex} 

\usepackage{enumitem}
\setlist{nosep,leftmargin=*}

\setlist[itemize]{leftmargin=*, itemsep=0.25ex, topsep=0.6ex, parsep=0ex}
\setlist[enumerate]{leftmargin=*, itemsep=0.25ex, topsep=0.6ex, parsep=0ex}

\title{Flow of Knowledge: Federated Fine-Tuning of LLMs in Healthcare under Non-IID Conditions}
\name{Zeyu Chen$^{\dagger}$, Yun Ji$^{\dagger}$, Bowen Wang, Liwen Shi, Zijie Zeng, Sheng Zhang$^{*}$}
\address{Graduate School in Shenzhen, Tsinghua University, Shenzhen 518055, China\\
Email: \{zy-chen23, ji-y20, wangbw23, liwenshi, zijiezeng\}@mails.tsinghua.edu.cn;\\ 
shengzhang@tsinghua.edu.cn\\
$^{\dagger}$ These authors contributed equally to this work.\quad
$^{*}$ Corresponding author.}

\begin{document}
\ninept
\maketitle
\begin{abstract}
Large language models (LLMs) show great promise in healthcare, but their applications are hindered by data privacy restrictions and the challenges of cross-institution collaboration. Sensitive medical data cannot be centralized, while non-independent and identically distributed (non-IID) characteristics across institutions further complicate convergence and fairness. To address these issues, we present a federated fine-tuning approach based on Low-Rank Adaptation (LoRA), enabling privacy-preserving knowledge flow across institutions. The method iteratively combines local LoRA adaptation with global parameter aggregation, allowing efficient knowledge sharing without exposing raw data. A blockchain identity scheme is used for identifying individual LLM in such a distributed network. We evaluate this approach on heterogeneous and highly non-IID medical text datasets, where experiments demonstrate that federated LoRA not only enhances cross-client generalization but also improves the performance of the weakest client, achieving stable convergence and fairer outcomes. These findings highlight federated LoRA fine-tuning as a practical and effective paradigm for adapting LLMs in healthcare, offering a new path for multi-center medical AI collaboration.
\end{abstract}
\section{Introduction}
\label{sec:intro}
Large Language Models (LLMs) have shown strong potential in vertical domains such as healthcare, finance, and law~\cite{singhal2025toward,liu2023fingpt,cui2023chatlaw}. While large-scale pretraining equips them with broad knowledge and reasoning abilities, domain-specific deployment still requires fine-tuning on specialized data~\cite{hagos2024recent,joshua2025domain}. However, conventional centralized training, which aggregates sensitive data from multiple institutions, poses serious privacy risks and conflicts with regulations such as HIPAA and GDPR~\cite{info15110697,sarwar2025fedmentalcare,truong2021privacy}. In highly regulated domains like healthcare, where data are fragmented across institutions with distinct specializations, such challenges make centralized fine-tuning impractical for cross-institutional collaboration.

Federated Learning (FL) has emerged as a promising paradigm to address this contradiction~\cite{wu2025survey}. By exchanging model parameters instead of raw data, FL enables distributed collaborative training while preserving privacy. However, in the context of LLM fine-tuning for vertical domains, non-independent and identically distributed (non-IID) data can lead to degraded global performance. 

Recent studies have applied FL to healthcare tasks such as biomedical information extraction~\cite{peng2024biomedical}, electronic health record (EHR) analysis, and mental health monitoring~\cite{sarwar2025fedmentalcare}, demonstrating feasibility under privacy constraints. 
There are also emerging attempts to combine FL with parameter-efficient tuning, for example through adapter modules or LoRA, to reduce communication overhead and improve scalability. 
However, most of these works focus on relatively homogeneous data distributions, smaller backbones, or narrow tasks (e.g., classification, NER), and lack systematic evaluation on multilingual, cross-format medical QA benchmarks.

To this end, we study federated fine-tuning of large language models in healthcare using parameter-efficient methods such as LoRA.To our best knowledge, federated fine-tuning of LLMs in healthcare under extreme non-IID conditions has not been systematically studied.


\noindent\textbf{Contributions.} The main contributions of this work are as follows:
\begin{itemize}

    \item \textbf{System implementation.} We provide the first systematic study of federated fine-tuning of LLMs in healthcare under extreme non-IID conditions. We implement a deployable multi-machine system, PrisMesh, built on an open-source Ray framework. The system supports LoRA-based local fine-tuning with frozen backbones and efficient adapter-only aggregation, ensuring low communication cost, scalability across heterogeneous GPU clusters, and practical readiness for cross-institution healthcare applications.

    \item \textbf{Heterogeneous setup.} To stress-test the federated fine-tuning process, we construct a challenging data distribution using multiple medical datasets that differ in language (Chinese vs. English), task format (licensing exams vs. entrance exams), and sample size imbalance.

    \item \textbf{Empirical findings.}  Through experiments on multiple medical QA datasets and LLMs of different sizes and architectures, we validate the effectiveness and scalability of the framework. Our experiments reveal that federated LoRA significantly improves fairness and accuracy across clients.

\end{itemize}

Additionally, this work includes Cypherium, a next-generation Layer-1 blockchain to provide transparent incentive schemes of decentralized training workload for cross-institutional collaboration in this work, with functions such as rewards and auditability. The blockchain scheme with unique hybrid proof-of-work and HotStuff Byzantine fault tolerance consensus is used to serve as verifiable identifiers for large models, ensuring fair rewards for parameter contributors with instant transaction settlement and absolute finality, and support secure data exchange and trustworthy intelligent computing networks in federated healthcare~\cite{wu2025blockchain}.

\section{Proposed Methods}
\label{sec:method}

\subsection{Federated Fine-Tuning Workflow}
\label{ssec:overview}
To address the challenges of non-IID data and strict privacy constraints in multi-institutional collaboration, we propose a federated fine-tuning framework for LLMs that supports deployed multi-machine training. The framework extends Low-Rank Adaptation (LoRA) into the federated setting, enabling collaborative training and knowledge fusion across heterogeneous clients while preserving local data privacy.

\begin{figure}[t]
  \centering
  \includegraphics[width=\linewidth]{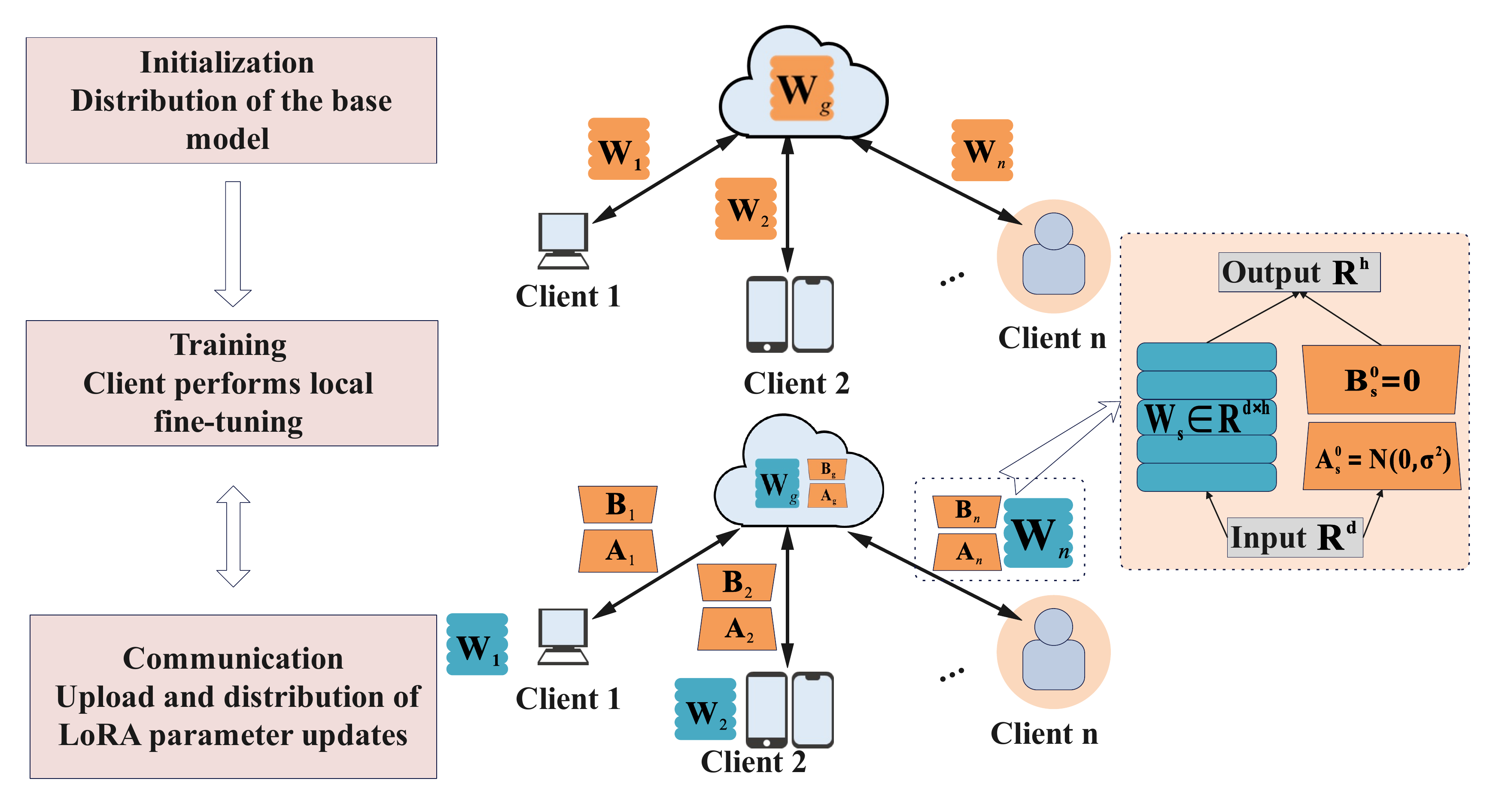}
  \caption{Overview of our federated fine-tuning framework.}
  \label{fig:framework}
\end{figure}

The framework consists of two types of nodes: a central aggregator node and multiple distributed client nodes. 
Each client is assigned a Cypherium public–private key pair, which is registered on the blockchain as a unique identifier. This guarantees data integrity and authenticity across the distributed network. Clients earn rewards in Cypherium tokens upon successfully completing their assigned workloads. 
Each client loads its private dataset, performs local LoRA-based fine-tuning with frozen backbone weights, and uploads only the low-rank adapter updates. The aggregator collects these updates, performs parameter aggregation (e.g., FedAvg), and redistributes the updated global LoRA adapters to all clients for the next round. The entire system is implemented on the Ray distributed framework, which abstracts client and server behaviors as actors, while handling communication, scheduling, and fault tolerance automatically, significantly reducing engineering complexity. This iterative process continues until convergence. The overall architecture and workflow are illustrated in Fig.~\ref{fig:framework}.

\begin{algorithm}[ht]
\caption{Federated LoRA Training Workflow}
\label{alg:fedlora}
\begin{algorithmic}[1]
\State \textbf{Initialization:} Aggregator distributes base model $W$ to all clients
\For{each round $t = 1,2,\dots,T$}
  \State \textbf{Parameter distribution:} Aggregator broadcasts $\Delta W^{(t)}$
  \For{each client $k = 1,2,\dots,K$ in parallel}
    \State Load local dataset $\mathcal{D}_k$
    \State Fine-tune base model with LoRA on $\mathcal{D}_k$
    \State Upload updated parameters $\Delta W_k^{(t)}$ to aggregator
  \EndFor
  \State \textbf{Aggregation update:} Apply FedAvg
  \[
    \Delta W^{(t+1)} \gets \tfrac{1}{K}\sum_{k=1}^K \Delta W_k^{(t)}
  \]
  \State \textbf{Synchronization:} Redistribute $\Delta W^{(t+1)}$ to all clients
\EndFor
\end{algorithmic}
\end{algorithm}

\subsection{Base Models and Low-Rank Adaptation (LoRA)}
\label{ssec:lora}

\subsubsection{Base model selection}
For multilingual medical QA tasks, we adopt Qwen3-1.8B/7B and LLaMA-3.1-8B as base LLMs~\cite{yang2025qwen3,grattafiori2024llama}. These models not only provide strong general-purpose language understanding and generation, but also incorporate extensive multilingual pretraining, forming rich cross-lingual priors. Such properties make them particularly suitable for federated scenarios involving cross-lingual and cross-institutional heterogeneity, while providing a stable initialization for LoRA-based fine-tuning.

\subsubsection{LoRA principle and parameter efficiency}
LoRA is a parameter-efficient fine-tuning approach. For a pretrained weight matrix $W \in \mathbb{R}^{d_{out}\times d_{in}}$, instead of updating $W$ directly, LoRA introduces a low-rank decomposition of the update:
\begin{equation}
h = Wx + \Delta Wx = Wx + \alpha \cdot BAx,
\end{equation}
where $A \in \mathbb{R}^{r \times d_{in}}$ and $B \in \mathbb{R}^{d_{out} \times r}$ are trainable low-rank matrices, $r$ is the rank, and $\alpha$ is a scaling factor. In the federated setting, only $(A,B)$ need to be trained and uploaded, dramatically reducing storage and communication costs.

\subsection{Federated Aggregation}
\label{ssec:agg}

We adopt the Federated Averaging (FedAvg) algorithm as the core aggregation mechanism. Unlike conventional FedAvg that aggregates full model weights, here we aggregate only LoRA adapters, reducing communication overhead and improving stability under non-IID conditions.

Let $K$ clients participate in round $t$, with client $k$ holding $n_k$ samples and total $N=\sum_{k=1}^{K} n_k$. The global LoRA update is:
\begin{equation}
\Delta W^{(t+1)} = \sum_{k=1}^{K} \frac{n_k}{N} \, \Delta W_k^{(t)},
\end{equation}
where $\Delta W_k^{(t)} = B_k^{(t)} A_k^{(t)}$ is the low-rank update from client $k$. In practice, the server receives $(A_k,B_k)$ from each client, reconstructs $\Delta W_k$, and performs the weighted averaging.

This adapter-only aggregation strategy yields two key benefits: (i) substantial reduction in communication and storage costs, making it suitable for multi-machine deployments with large numbers of clients; and (ii) improved stability, as the low-rank updates serve as a soft constraint on parameter space, particularly beneficial under cross-institution non-IID data distributions.

\section{Experiments and Results}
\label{sec:exp}

\subsection{Federated Dataset Setup}

To evaluate the effectiveness of the proposed framework in real-world medical scenarios, we construct a federated learning environment with extremely heterogeneous distributions on three clients, which can be demonstrated as follows :
\begin{itemize}
    \item \textbf{MedQA-CN:} Chinese medical licensing examination dataset, covering core subjects such as internal medicine and surgery, with 27,400 training and 3,426 test questions.
    \item \textbf{MedQA-EN:} English USMLE dataset, including Step~1/2/3 questions, with 10,178 training and 1,273 test questions.
    \item \textbf{MedMCQA-IN:} Indian medical entrance examination dataset, including AIIMS and NEET exams, with 33,464 training and 4,183 test questions.
\end{itemize}

This setup presents a typical extreme non-IID scenario:
\begin{itemize}
    \item \textbf{Language heterogeneity:} Chinese and English corpora with distinct terminology and expression styles.
    \item \textbf{Exam format divergence:} Licensing, certification, and entrance exams focusing on different knowledge areas and question types.
    \item \textbf{Data imbalance:} The largest client (MedMCQA-IN) contains 3.3$\times$ more samples than the smallest client (MedQA-EN).
\end{itemize}

\noindent
The goal of our experiments is to collaboratively optimize a shared large language model 
via federated learning, enhancing performance across all medical domains while preserving 
data privacy, with particular emphasis on improving the weakest-performing client. Moreover, integration of the Cypherium blockchain enhances communication and trust within the distributed network.while rewarding clients for their contributions of computing resources and data—effectively establishing a proof-of-useful-work mechanism.

\subsection{Experimental Setup}
\label{sec:exp_setup}
\subsubsection{Base Experimental Setup}

We conduct federated training with three communication rounds, with each client running one local epoch per round. LoRA uses rank $r=8$, scaling $\alpha=32$, and dropout 0.1. Experiments are deployed on a heterogeneous GPU cluster and repeated on multiple model scales (Qwen3-1.7B, Qwen3-8B, LLaMA-3.1-8B) under identical hyperparameters for fair comparison. 
The complete configuration is summarized in Table~\ref{tab:exp-setup}.

\begin{table}[ht]
\centering
\caption{Training setup, LoRA config, platform, and models}
\label{tab:exp-setup}
\begin{tabular}{ll}
\hline
\textbf{Training} & Communication rounds: 3; Local epochs: 1 \\
& Optimizer: AdamW ($\beta_1=0.9,\ \beta_2=0.999$) \\
& Learning rate: $3\times 10^{-5}$ (linear decay) \\
& Batch size: 1 (accumulation: 4) \\
\hline
\textbf{LoRA} & Rank $r=8$; Scaling $\alpha=32$; Dropout=0.1 \\
& Injected layers: q/k/v/o projections; gate/up/down MLPs \\
\hline
\textbf{Platform} & Hardware: Metax C500, N260, NVIDIA A100 GPUs \\
& Software: PyTorch 2.0.1, Transformers 4.45.0, Ray 2.23.0 \\
\hline
\textbf{Models} & Qwen3-1.7B, Qwen3-8B, LLaMA-3.1-8B \\
\hline
\end{tabular}
\end{table}

\subsubsection{Benchmarks and Evaluation}
We compare the following approaches:
\begin{itemize}
  \item \textbf{Base Model:} pretrained backbone LLMs without any fine-tuning.
  \item \textbf{Single-client LoRA:} LoRA fine-tuning on a single client dataset.
  \item \textbf{Federated LoRA:} LoRA fine-tuning under federated aggregation with FedAvg.
\end{itemize}

In addition to per-dataset accuracy, we report three aggregate metrics 
to capture both overall performance and fairness in federated settings. 
\textbf{Macro-Acc} is the arithmetic mean of per-client accuracies,
\begin{equation}
\label{eq:macroacc}
\mathit{Macro\text{-}Acc} = \frac{1}{K} \sum_{k=1}^K acc_k,
\end{equation}
reflecting average performance across clients. 
\textbf{Minimum Accuracy (Min-Acc)} is defined as
\begin{equation}
\label{eq:minacc}
\mathit{Min\text{-}Acc} = \min_{k} acc_k,
\end{equation}
highlighting the weakest client’s performance under heterogeneous 
distributions. Finally, \textbf{Harmonic Mean (H-mean)}, given in 
Eq.~\ref{eq:hmean}, penalizes low-performing clients more heavily 
than Macro-Acc, providing a stricter fairness measure 
\cite{wu2025survey}. 

Together with per-dataset accuracy, these metrics allow us to evaluate 
both the effectiveness and fairness of federated LoRA training.

Specifically, we report:
\begin{itemize}
  \item \textbf{Per-dataset Accuracy:} accuracy on each client’s test set.
  \item \textbf{Macro-Acc:} arithmetic mean of accuracies across clients (Eq.~\ref{eq:macroacc}).
  \item \textbf{Min-Acc:} the lowest accuracy among clients (Eq.~\ref{eq:minacc}).
  \item \textbf{H-mean:} harmonic mean of accuracies across clients,
\begin{equation}
\label{eq:hmean}
\mathit{H\text{-}mean} = \frac{K}{\sum_{k=1}^K \tfrac{1}{acc_k}}.
\end{equation}
\end{itemize}

\subsubsection{Performance and Comparison}

\begin{enumerate}[label=\Alph*.]

    \item \textbf{Cross-client transfer of Single-client LoRA.} 
    LoRA fine-tuning on a single client not only improves accuracy on the 
    corresponding local dataset but also transfers positively to other clients. 
    For instance, fine-tuning solely on MedQA-EN improves accuracy on 
    MedQA-CN and MedMCQA-IN by +5.2\% and +3.8\% over the base model, 
    demonstrating that LoRA adapters capture generalizable medical knowledge 
    beyond a single dataset.

    Table~\ref{tab:single_client} presents the cross-dataset evaluation results of single-client LoRA fine-tuning on Qwen3-1.7B. Each LoRA model improves its own dataset while also yielding gains on unseen clients, benefiting from the backbone’s multilingual pretraining that provides rich cross-lingual priors and stable initialization. This shows that LoRA captures transferable knowledge rather than overfitting to a single domain.

    \begin{table}[ht]
    \centering
    \caption{Single-client LoRA fine-tuning on Qwen3-1.7B: cross-dataset evaluation (Accuracy \%).}
    \label{tab:single_client}
    \resizebox{\linewidth}{!}{
    \begin{tabular}{lccc}
    \hline
    Model & MedQA-CN & MedQA-EN & MedMCQA-IN \\
    \hline
    Baseline (Qwen3-1.7B) & 58.46 & 41.01 & 34.02 \\
    + LoRA (Client 1: IN) & 59.60 & 40.06 & 46.59 \\
    + LoRA (Client 2: CN) & 68.24 & 44.93 & 44.68 \\
    + LoRA (Client 3: EN) & 63.11 & 48.39 & 44.20 \\
    \hline
    \end{tabular}}
    \end{table}

    \item \textbf{Fairness improvement with Federated LoRA.} 
    Compared with single-client fine-tuning, federated aggregation enhances 
    performance balance across clients. Specifically, Table~\ref{tab:results_1.7b} 
    shows that min-acc improves by 2.3--4.1\% over the best single-client LoRA, H-mean increases by 1.8--3.2\%, and macro-accuracy remains stable without degradation. This indicates that aggregation mitigates the weakest-client performance drop while preserving overall accuracy. 

    In addition, Figure~\ref{fig:bar_8b} illustrates the same trend on Qwen3-8B, where Federated LoRA achieves more balanced improvements across datasets compared to the strongest single-client LoRA, further confirming the fairness advantage of aggregation.

    \begin{table}[ht]
    \centering
    \caption{Results on Qwen3-1.7B across three datasets. Federated LoRA improves minimum accuracy and H-mean.}
    \label{tab:results_1.7b}
    \resizebox{\linewidth}{!}{
    \begin{tabular}{lccc}
    \hline
    \textbf{Model} & \textbf{Macro-Acc} & \textbf{Min-Acc} & \textbf{H-mean} \\
    \hline
    Baseline (1.7B)            & 44.50 & 34.02 & 42.32 \\
    Best Single LoRA (CN)      & 52.44 & 44.68 & 50.45 \\
    Federated LoRA (Merged\_2) & \textbf{53.56} & \textbf{46.40} & \textbf{52.03} \\
    \hline
    \end{tabular}}
    \end{table}

    \begin{figure}[ht]
    \centering
    \includegraphics[width=\linewidth]{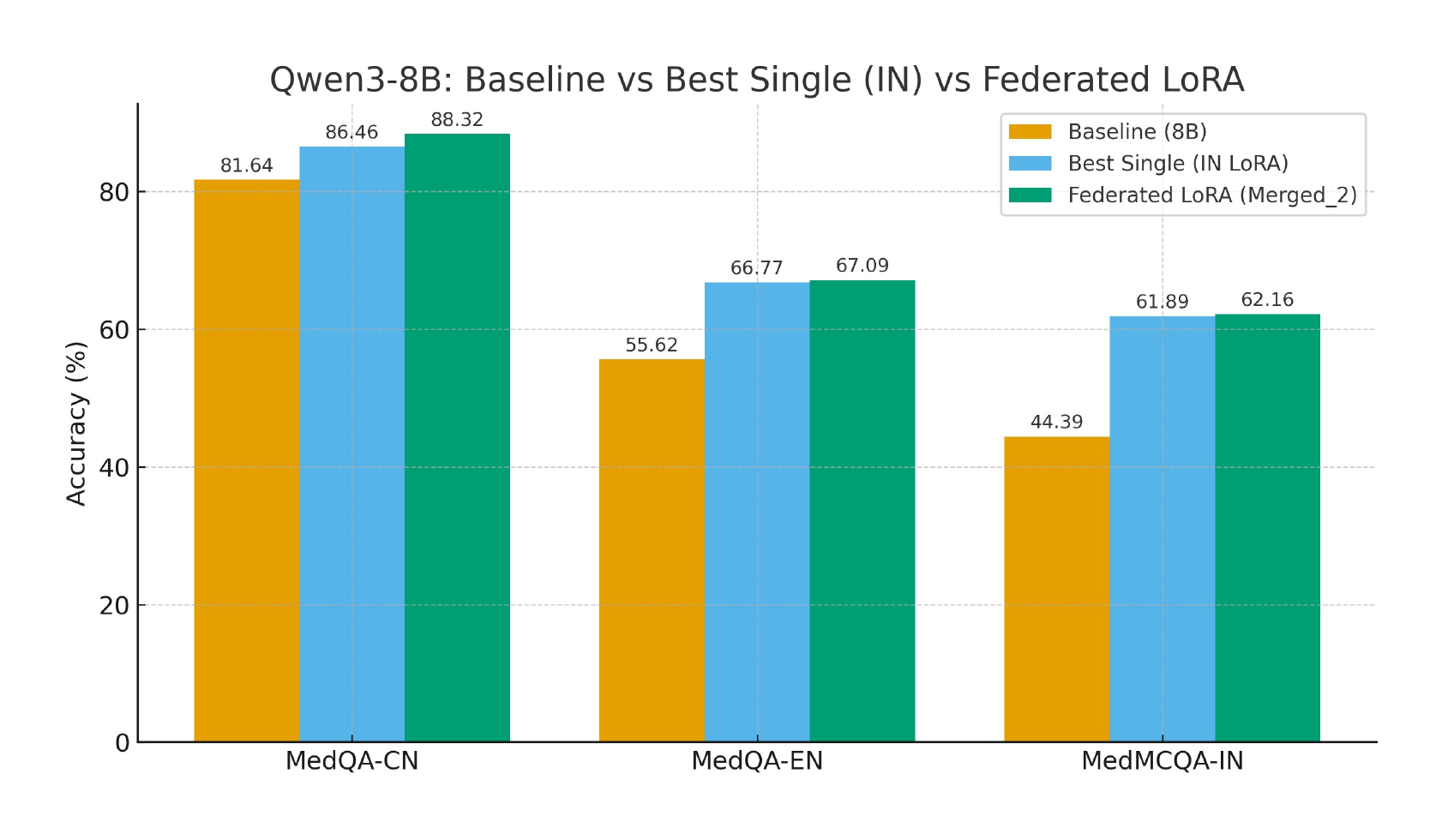}
    \caption{Accuracy comparison on Qwen3-8B: Baseline vs Best Single vs Federated LoRA.}
    \label{fig:bar_8b}
    \end{figure}

    \item \textbf{Cross-model consistency.} 
    Experiments on Qwen3-1.7B, Qwen3-8B, and LLaMA-3.1-8B reveal a consistent trend: single-client LoRA improves not only local datasets but also transfers across clients, while Federated LoRA further boosts min-acc and H-mean. As shown in Table~\ref{tab:results_8b_llama}, these gains also scale with model size, demonstrating that aggregation benefits are robust across architectures. The backbone models’ extensive multilingual pretraining provides stable cross-lingual priors, which complements federated aggregation and makes LoRA adaptation broadly effective under heterogeneous settings.

    \begin{table}[ht]
    \centering
    \caption{Accuracy of other models across three datasets (\%). 
    }
    \label{tab:results_8b_llama}
    \resizebox{\linewidth}{!}{
    \begin{tabular}{lccc}
    \hline
    \multicolumn{4}{c}{\textbf{Qwen3-8B}} \\
    \hline
    Model & MedQA-CN & MedQA-EN & MedMCQA-IN \\
    \hline
    Base (8B)        & 81.64 & 55.62 & 44.39 \\
    LoRA (Client IN) & 86.46 & 66.77 & 61.89 \\
    LoRA (Client CN) & 89.40 & 65.04 & 60.75 \\
    LoRA (Client EN) & 86.31 & 69.21 & 59.79 \\
    Federated (Merged\_2) & 88.32 & 67.09 & 62.16 \\
    \hline
    \multicolumn{4}{c}{\textbf{LLaMA-3.1-8B}} \\
    \hline
    Model & MedQA-CN & MedQA-EN & MedMCQA-IN \\
    \hline
    Base (8B)        & 64.24 & 55.85 & 44.42 \\
    LoRA (Client IN) & 70.78 & 60.80 & 56.37 \\
    LoRA (Client CN) & 71.92 & 59.39 & 53.10 \\
    LoRA (Client EN) & 69.79 & 60.80 & 49.96 \\
    Federated (Merged\_2) & 72.80 & 60.57 & 55.15 \\
    \hline
    \end{tabular}}
    \end{table}

\end{enumerate}
\subsection{Multi-round Aggregation Dynamics}
\begin{figure}[ht]
\centering
\includegraphics[width=\linewidth]{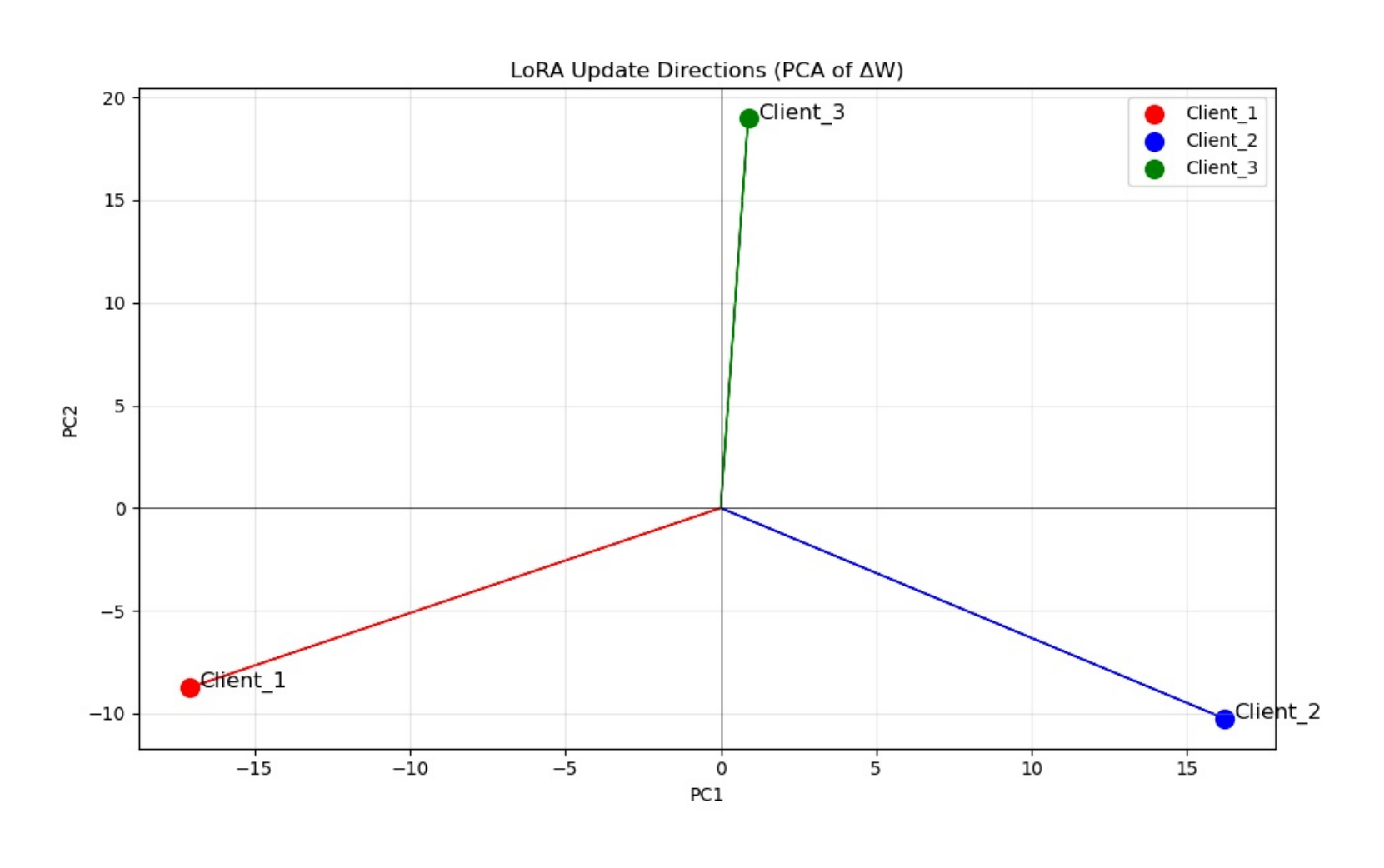}
\caption{PCA visualization of LoRA parameter updates from three clients in the first communication round}
\label{fig:pca}
\end{figure}

During multi-round federated aggregation, we observe a critical phenomenon: after the first communication round, the aggregated global model may temporarily underperform the best single-client LoRA model. This can be attributed to the divergence of local update directions under heterogeneous data distributions. As shown in the PCA visualization (Fig.~\ref{fig:pca}), LoRA updates from different clients are widely dispersed in parameter space, reflecting distinct feature patterns and decision boundaries. The initial aggregation represents a compromise across divergent client updates, which reduces alignment with any single client distribution and causes a temporary performance drop.

As additional rounds proceed, the global model provides clients with a more consistent initialization, reducing variance in update directions. This iterative process enables effective alignment across institutions, ultimately allowing the aggregated model to surpass single-client LoRA models. LoRA’s low-rank constraint further acts as a regularizer that smooths update trajectories, while strong multilingual priors in the base models facilitate cross-lingual knowledge alignment. Together, these dynamics demonstrate how federated fine-tuning serves as a mechanism of knowledge flow, where complementary expertise from different institutions is gradually integrated, ultimately surpassing single-client LoRA performance.

\section{Future Directions}
\label{sec:disc}

Beyond the current scope, several promising directions remain to be explored. 
First, extending federated fine-tuning beyond medical QA to clinical text generation, multimodal reasoning with imaging data, and decision-support dialogue systems could further validate the generality of the framework. 
Second, more advanced aggregation strategies (e.g., FedProx, SCAFFOLD, or personalized FL approaches) may mitigate non-IID challenges more effectively and enable finer-grained adaptation to institutional heterogeneity. 

The incorporation of blockchain-based identity and incentive mechanisms provides another important avenue. 
Each client can be assigned a Cypherium public–private key pair, which is registered on the blockchain as a unique identifier. 
This guarantees data integrity and authenticity across the distributed network. Furthermore, in-proposed blockchain-based scalable architecture compatible with Cypherium network, provides a promising future direction to encourage sustained participation and ensure accountability in federated healthcare ecosystems, making it a powerful foundation for future LLM intelligent computing networks.

\section{Conclusion}
\label{sec:concl}

In this work, we investigated federated fine-tuning of large language models in healthcare under extreme non-IID conditions. We show that knowledge acquired through local LoRA adaptation can flow across institutions via federated aggregation, improving fairness by boosting the weakest client while preserving strong ones, and ensuring stable convergence despite heterogeneous data. These findings highlight federated LoRA as a practical paradigm for equitable and privacy-preserving collaboration across healthcare institutions.The integration of the Cypherium blockchain showcases a promising convergence of AI and blockchain technologies, fostering fairness and collaboration among distributed clients. While our experiments focus on medical QA tasks, the approach is easily transfered to other applications.

%

\bibliographystyle{IEEEbib}
\bibliography{refs}

\end{document}